\documentclass[11pt]{article}
\usepackage{graphicx}
\usepackage{bbm,bm}

\newcommand{\hatp}{\hat{\bm p}}
\newcommand{\hatk}{\hat{\bm k}}
\newcommand{\hatn}{\hat{\bm n}}
\newcommand{\hatl}{\hat{\bm l}}

\begin{document}


\begin{center} 
{\Large\bf   Testing of $P$ and $CP$ Symmetries with $e^+e^- \to J/\psi \to \Lambda\bar \Lambda$   }
\par\vskip20pt
 X.G. He$^{1,2,3}$ and J.P. Ma$^{4,5,6}$,    \\
{\small {\it
$^1$ Tsung-Dao Lee Institute \& School of Physics and Astronomy,
Shanghai Jiao Tong University, Shanghai 200240, China \\
$^2$Key Laboratory for Particle Astrophysics and Cosmology (MOE)
{\rm \&} Shanghai Key Laboratory for Particle Physics and Cosmology,
Shanghai Jiao Tong University, Shanghai 200240, China\\
$^3$ Department of Physics and National Center for Theoretical Sciences, National Taiwan University, Taipei 10617\\
$^4$ CAS Key Laboratory of Theoretical Physics, Institute of Theoretical Physics, P.O. Box 2735, Chinese Academy of Sciences, Beijing 100190, China\\
$^5$ School of Physical Sciences, University of Chinese Academy of Sciences, Beijing 100049, China\\
$^6$ School of Physics and Center for High-Energy Physics, Peking University, Beijing 100871, China}}
\end{center}

\vskip 1cm
\begin{abstract}
 
We propose to test $P$- and $CP$-symmetries with the process $e^+e^- \to J/\psi \to \Lambda\bar \Lambda$. 
The general form of the angular distribution of the process is derived and several observables for the proposed test are introduced. With these observables one can distinguish the effects of $P$-violations  in the production of $J/\psi$ 
from  its decay into a pair of $\Lambda \bar \Lambda$. 
Numerical estimations for proposed asymmetries of $P$-violation due to $Z$ exchange in SM are given. Our results show that
BESIII has reached the sensitivity to probe these effects.  At the proposed super-tau-charm factory with much enhanced luminosity it will be able to test the SM predictions with high precision. $CP$ violating effect in $J/\psi \to \Lambda \bar \Lambda$ can also be tested. In particular, BESIII data already reached the sensitivity to prove $CP$ violating effect due to current upper limit of electric dipole moment of $\Lambda$.  At the proposed super-tau-charm factory the upper limit can be improved by an order of magnitude. 

\end{abstract}      
\vskip 5mm

\vskip40pt

\noindent
{\bf 1. Introduction}

\par\vskip5pt
The symmetries of Parity ($P$), Charge conjugation ($C$) and Time-reversal ($T$), are fundamental space-time discrete symmetries. Their conservation and violation play an important role in understanding the laws of Nature.  The $P$- and $CP$-violation have been already found in several places and can be described in the framework of the standard model (SM)\cite{PDG}. It is important to further test these symmetries in all possible ways to probe new physics beyond SM. Several facilities are actively carrying out such precision tests with large data samples.  In this letter we focus on possible tests of $P$- and $CP$-symmetry using the resonant production of $J/\psi$ at a $e^+e^-$ collider which subsequently decays into a pair of $\Lambda \bar \Lambda$.

Tests of the discrete symmetries require large data samples in experiment. There are large data samples of $e^+e^-\to J/\psi \to \Lambda\bar\Lambda$ collected by BESIII at BEPCII. 
At Super-Tau-Charm Factory (STCF) proposed in \cite{STCFC,STCFR} the data samples will be more 
larger than those collected by BESIII.  Data from these experiments can provide much needed information. This process has been studied in BESIII experiment, but for testing discrete symmetries only for $CP$-violation in the decays of the produced $\Lambda$ and $\bar\Lambda$ \cite{BESIII:2022qax,BESIII:2018cnd,BESIII:2021ypr}.
 There are no tests with the production of $J/\psi$ and its decay  into $\Lambda\bar\Lambda$. 
 In general  the violation of discrete symmetries in the chain process can also happen in these two places.

It is noted that without observation of polarizations of the final- and initial state one can        
not test discrete symmetry with  $e^+e^-\to J/\psi \to \Lambda\bar\Lambda$\cite{CPZ}. Thanks to the weak decays of $\Lambda$ and $\bar 
\Lambda$, the polarization of the produced $\Lambda$ and $\bar\Lambda$ can be observed\cite{LeeYang}. 
This makes the tests of symmetries possible in $e^+e^- \to J/\psi \to \Lambda\bar\Lambda$, where the polarization of the initial state is not observed.   
$CP$-symmetry test in the decay of $J/\psi\to \Lambda\bar\Lambda$ has been studied in \cite{HMM}. However, the results only apply for the unpolarized $J/\psi$. In the process considered here, $J/\psi$ is polarized.
We propose several observables for the symmetry test without observing the polarization of the initial state. Using these observables 
 effects due to $P$-violation  in  $e^+e^- \to J/\psi$ and in $J/\psi\to \Lambda \bar \Lambda$ can be separately measured. $CP$-violation appearing in $J/\psi\to \Lambda \bar \Lambda$ can also be detected with our observables. The proposed observables can already be measured with the data sample 
 at BES. Using polarized beams to study $P$-violation in the process for measuring Weinberg angle has been studied in \cite{BGR}.

In the following we will first give the most general form 
of the angular distribution only based on covariance and discuss the constraints of discrete symmetries in Sect.2.  In Sect.3 we give each component in the angular distribution with a general parametrization of 
the amplitude. In Sect.4 Observables and corresponding asymmetries for testing discrete symmetries are suggested and their results 
are given.  Numerical results and discussions are provided in Sect.5. Finally we will give our summary in Sect.6.

\par\vskip10pt

\noindent 
{\bf 2. The  General Angular Distributions and Its Constraints from Symmetries} 

\par\vskip5pt 
We consider the production of $J/\psi$ through the annihilation of an $e^+e^-$ pair and its sequential decay into a $\Lambda \bar\Lambda$ pair:  
\begin{equation} 
   e^- (p_1) + e^+ (p_2)\to J/\psi   \to \Lambda (k_1,s_1) + \bar \Lambda (k_2,s_2),   
\end{equation} 
where the spins of the initial state are averaged.  Momenta and spins of $\Lambda$ and $\bar \Lambda$ are indicated in the brackets. In the center of mass frame (CMF), the momenta are
\begin{equation} 
   p_1^\mu = (E_c, {\bm p} ), \quad  p_2^\mu = (E_c, -{\bm p} ), \quad k_1^\mu = (k^0, {\bm k} ), \quad k_2^\mu = (k^0, -{\bm k} ). 
\end{equation} 
In this frame the $J/\psi$ is at rest and $E_c$ is the half of the $J/\psi$-mass $M_{J/\psi}$. We define the spin vector $s_1$ of $\Lambda$ as the spin vector in its rest frame which is obtained  through a Lorentz boost without any spacial rotation from the frame of CMF. Hence, the spin vector 
has only nonzero spacial components, i.e., $s_1^\mu = (0, {\bm s}_1)$. The spin vector $s_2$ of $\bar\Lambda$ is defined similarly.  For convenience we introduce three unit vectors and the variable $\omega$:
\begin{equation} 
   \hat{\bm p}   =\frac{{\bm p} }{\vert {\bm p} \vert} , \quad \hatk =\frac{{\bm k}}{\vert {\bm k} \vert}, \quad
     \hatn  =\frac{ {\bm p} \times {\bm k}} {\vert {\bm p} \times {\bm k}  \vert }, \quad 
      \omega =\hatp\cdot\hatk,  
\end{equation}  
where the first two vectors are for the momentum direction of $e^-$ and $\Lambda$, respectively. $\hatn$ is the direction normal to the production plane.    
 
\par 
For the $J/\psi$ decay we can write the decay amplitude ${\mathcal T}$  and the density matrix $R$ for the decay as:
\begin{equation} 
      {\mathcal T} = \epsilon^\mu  {\mathcal A}_\mu, \quad   R(\hatp, \hatk, {\bm s}_1, {\bm s}_2)  = {\mathcal T}  {\mathcal T} ^\dagger =\rho^{ij}  {\mathcal M}^{ij}, 
 \end{equation} 
 with 
 \begin{equation}      
     {\mathcal M}^{ij} =    {\mathcal A}^i {\mathcal A}^{*j}, \quad 
      \rho^{ij} =  \epsilon^i \epsilon^{*j}.    
\label{RHO}       
\end{equation} 
where $\epsilon^\mu$ is the polarization vector of $J/\psi$ in its rest frame with $\epsilon^0=0$.   $\rho^{ij}$ is the density matrix for the production of $J/\psi$,  ${\mathcal M}$ is the density matrix of the decay. It is clear that ${\mathcal M}$ only depend
on ${\bm k}$ and ${\bm s}_{1,2}$.  $\rho^{ij}$ contains the information about the production. In our case
that $J/\psi$ is produced though $e^+e^-$ annihilation, it only 
depends on ${\bm p}$. With rotational covariance, $\rho$ can be decomposed as
\begin{equation} 
    \rho^{ij}(\hatp) =\frac{1}{3}\delta^{ij} -i d_J \epsilon^{ijk} \hat p^k -\frac{c_J}{2} \biggr ( \hat p^i \hat p^j -\frac{1}{3} \delta^{ij} \biggr ),
\label{RHOC}  
\end{equation} 
where we have normalized $\rho$ so that ${\rm Tr} \rho =1$. $d_J$ and $c_J$ are real constants. 
Parity conservation implies that there is no asymmetric part in $\rho$, i.e., $d_J=0$.  $CP$-symmetry gives no constraint.  If the annihilation is into one virtual photon in the first step and the photon is then converting to $J/\psi$,   we have $c_J=1$.
The density matrix $R$ contains all information about the considered process. 

The decomposition of  ${\mathcal M}^{ij}$ with the three vector $\hatk$ and ${\bm s}_{1,2}$ is complicated. However, the spin-dependence of ${\mathcal M}$ or $R$ can be easily found as: 
\begin{equation} 
R\left (\hatp, \hatk, {\bm s}_1, {\bm s}_2 \right )  = a (\omega ) +  {\bm s}_1\cdot {\bm B}_1 (\hatp, \hatk)
   +   {\bm s}_2\cdot {\bm B}_2 (\hatp, \hatk)  + s_1^i s_2^j C^{ij} (\hatp, \hatk). 
\end{equation} 
With rotational covariance ${\bm B}_{1,2}$ and $C^{ij}$ can be decomposed as:   
\begin{eqnarray} 
    {\bm B}_1  (\hatp, \hatk) &=& \hatp b_{1p} (\omega) + \hatk b_{1k} (\omega)  
       + \hatn b_{1n} (\omega), 
\quad
    {\bm B}_2  (\hatp, \hatk) = \hatp  b_{2p} (\omega) + \hatk b_{2k} (\omega)  
       + \hatn b_{2n} (\omega), 
\nonumber\\
   C^{ij} (\hatp, \hatk) &=& \delta^{ij} c_0 (\omega) +\epsilon^{ijk} 
      \biggr (\hat p^k c_1 (\omega) + \hat k^k c_2 (\omega) +\hat n^k c_3 (\omega) \biggr ) 
\nonumber\\
    && +\left  (\hat p^i \hat p^j -\frac{1}{3}\delta^{ij} \right ) c_4 (\omega) + \left  (\hat k^i \hat k^j -\frac{1}{3}\delta^{ij} \right ) c_5 (\omega)+ \left  (\hat p^i \hat k^j + \hat k^i \hat p^j  -\frac{2}{3}\omega\delta^{ij}  \right ) c_6 (\omega)
\nonumber\\
    && +\left  (\hat p^i \hat n^j + \hat n^i \hat p^j  \right ) c_7 (\omega)
    +\left  (\hat k^i \hat n^j + \hat n^i \hat k^j  \right ) c_8 (\omega).  
\label{DECR}     
\end{eqnarray} 
Without constraints from $P$ and $CP$ symmetries, there are 16 scalar functions charactering the process.

The spin of $\Lambda$ or $\bar\Lambda$ can be measured with its weak decay $\Lambda \to p +\pi$ 
or $\bar \Lambda \to \bar p +\pi$, respectively. It is noted that the spin of $\Lambda$ or $\bar \Lambda$ measured through their decays are not exactly the same as those in the production.  This is due to Larmor precession in the magnetic field of the detector. 
The effect of Larmor precession is small\cite{LiMa}. We will neglect the effect.    The differential distributions of the decays in the rest frame can be written as:
\begin{equation} 
   \frac{ d\Gamma_{\Lambda}} { d\Omega_p } ({\bm s}_1, \hatl_p )  \propto 1 + \alpha {\bm  s}_1 \cdot \hatl_p,\quad 
   \frac{ d\Gamma_{\bar \Lambda}} {d\Omega_{\bar p} } ({\bm s}_2, \hatl_{\bar p} )  \propto 1 -\bar \alpha {\bm s}_2 \cdot \hatl_{\bar p}, 
\end{equation} 
where $\hatl_p$ and  $\hatl_{\bar p}$ is the direction of the momentum of the proton or anti-proton in the rest frame of $\Lambda$ or $\bar\Lambda$,  respectively.  $\Omega_{p,\bar p}$ is the corresponding solid angle. $\alpha$ and $\bar \alpha$ are constants. $CP$-conservation gives that $\alpha=\bar\alpha$. 
Taking the distributions and $R$ as density matrices, the spin vector ${\bm s}_{1,2} $ should be understood as Pauli matrix ${\bm\sigma}_{1,2}$, respectively.

With the above discussion, the general angular distribution for the process where the produced 
$\Lambda$ and $\bar\Lambda$ decays through weak interaction sequentially, is given by:
\begin{eqnarray} 
  \frac{d\sigma}{d\Omega_k d \Omega_p d\Omega_{\bar p} }  &\propto & {\mathcal W}(\Omega), 
\nonumber\\  
   {\mathcal W}(\Omega) & =& \frac{1}{4}   {\rm Tr} \biggr ( R (\hatp, \hatk, {\bm \sigma}_1, {\bm\sigma}_2) ( 1 + \alpha {\bm \sigma}_1 \cdot \hatl_p )
      (1 -\bar \alpha {\bm \sigma}_2 \cdot \hatl_{\bar p} ) \biggr ) 
\nonumber\\
     &=& a (\omega ) +   \alpha \hatl_p\cdot {\bm B}_1 (\hatp, \hatk)
    -\bar\alpha   {\hatl}_{\bar p}\cdot{\bm B}_2 (\hatp, \hatk)   -\alpha\bar\alpha \hat l_p^i  \hat l_{\bar p} ^j C^{ij} (\hatp, \hatk),  
\label{GAD}           
\end{eqnarray}                    
where $\Omega_k$ is the solid angle of $\hatk$.  We use $\Omega$ to denote the set of solid angles $\Omega_{k,p,\bar p}$ as variables of ${\mathcal W}$. With this general form and the decomposition in Eq.(\ref{DECR}) one can derive the general angular 
distribution in any coordinate system which is convenient for analyzing experimental data.  The angular distribution with the constraints of discrete symmetries have been given in \cite{BGR,FK} in a specified  coordinate system. 
\par
The general angular distribution is derived only by using rotational covariance. Discrete symmetries give  
constraints on it.  In the following we discuss on the constraints for $R$.  If  $C$-symmetry holds, we have:   
\begin{equation} 
R\left (\hatp, \hatk,  {\bm s}_1, {\bm s}_2 \right ) = R\left (-\hatp, -\hatk,  {\bm s}_2, {\bm s}_1 \right ) 
\end{equation} 
This implies for the functions in Eq.(\ref{DECR}):
\begin{equation} 
    b_{1m} (\omega) = -b_{2m} (\omega), \quad m=k, p. \quad   b_{1n} (\omega) = b_{2n} (\omega), \quad  
    c_3 (\omega)=c_7(\omega) = c_8(\omega)=0. 
\label{CSC}     
\end{equation}      
If parity is conserved, we have 
\begin{equation} 
R\left (\hatp, \hatk, {\bm s}_1, {\bm s}_2 \right ) =R\left (-\hatp, -\hatk, {\bm s}_1, {\bm s}_2 \right ) . 
\end{equation} 
This indicates that the following functions must be zero:
\begin{equation} 
    b_{1p} (\omega) = b_{2p} (\omega)  =b_{1k} (\omega) = b_{2k} (\omega) = c_1(\omega) = c_2 (\omega)=c_7(\omega) = c_8(\omega)=0. 
\label{PSC}     
\end{equation}     
Therefore, if $C$- and $P$ are conserved, $R$ is characterized only by 6 functions of $\omega$ instead of 
16 in Eq.(\ref{DECR}).  If CP-symmetry helds, we have the following relations:
 \begin{eqnarray} 
    b_{1m} (\omega) = b_{2m} (\omega), \quad m=p,k,n.  \quad c_i (\omega) =0, \quad i=1,2,3. 
\label{CPSC}     
\end{eqnarray} 
The symmetry of time-reversal does not give constrains as strong as those discussed in the above. It only indicates 
that the difference 
\begin{equation} 
R\left (\hatp, \hatk, {\bm s}_1, {\bm s}_2 \right ) -R\left (-\hatp, -\hatk, -{\bm  s}_1, -{\bm s}_2 \right )  
\end{equation}    
is proportional to the interference between absorptive- and dispersive part of the amplitude. 
The difference gives the so-called $T$-odd effects. E.g., it is easy to find that 
nonzero $b_{1n,2n}$ is from the interference and results in that 
the $\Lambda$ is polarized along the $\hatn$-direction as observed in experiment. 

With the constrains given in the above, tests of discrete symmetries of  $C$, $P$ and $CP$  are to check the relations in Eqs.(\ref{CSC}), (\ref{PSC}) and (\ref{CPSC}), respectively.

\par\vskip10pt

\noindent 
{\bf 3. The General  Form of the Decay Amplitude and Results for $R$} 

\par\vskip5pt

The general form of the amplitude of $J/\psi\to \Lambda\bar\Lambda$ can be written as:
\begin{eqnarray} 
  {\mathcal A}^\mu  =  \bar u(k_1) \biggr [ \gamma^\mu F_V +\frac{i}{2 m_\Lambda }\sigma^{\mu\nu} q_\nu H_\sigma + \gamma^\mu \gamma_5 F_A   +\sigma^{\mu\nu} q_\nu \gamma_5  H_{T} \biggr ] v (k_2), 
\label{AMP}   
\end{eqnarray} 
where $F_V, F_A, H_\sigma$ and $H_T$ are in general complex numbers, or form factors at fixed $q^2$
with $q^2=M^2_{J/\psi}$. $m_\Lambda$ is the mass of $\Lambda$. 
$P$ conservation implies that $F_A=0$. If $CP$ is conserved, we have $H_{T}=0$. 
Since it is expected that $F_A$ and $H_T$ will be much smaller than other two constants in the amplitude, 
we will only take 
the interference effect from the last two terms in Eq.(\ref{AMP}) into account.  
In calculating of $R$ we need the density matrix $\rho$ in Eq.(\ref{RHOC}). We assume that the dominant 
contribution for the production of $J/\psi$ is from one-photon exchange. With this assumption we take $c_J$ 
as $1$.  

In Eq.(\ref{AMP}) the form factor $F_V$ or $H_\sigma$ is the analogy to the Dirac- and Pauli form factor of a bayron, 
respectively.  In \cite{FK} different form factors are introduced. We find 
that with the form factors in \cite{FK} the results are simpler than those expressed with $F_V$ and $H_{\sigma}$. Hence, we introduce two form factors $G_{1,2}$ which are related to $F_V$ and $H_{\sigma}$ as:   
\begin{equation} 
    G_1 = F_V +H_\sigma, \quad  G_2 = G_1 - \frac{(k_1-k_2)^2}{2 m_\Lambda^2} H_\sigma,  \label{relation}
\end{equation}   
where $G_1 (G_2) $ is  proportional to  $G_M^\psi (G_E^\psi)$  defined in \cite{FK}. It is noted that 
$G_1 (G_2) $ is not an analogy to Sachs form factors. 

The $P$- and $C$- conserving parts of $R$ is given by the following nonzero functions: 
\begin{eqnarray} 
  a(\omega) &=& E_c^2 \biggr [ \vert G_1 \vert^2 + y_m^2 \vert G_2\vert^2 + \omega^2 ( \vert G_1 \vert^2 - y_m^2 \vert G_2\vert^2 )\biggr ], \quad 
\nonumber\\
     b_{1n} (\omega) &=& b_{2n}(\omega)  =  2\vert \hatp\times \hatk\vert  E_c^2  \omega y_m   {\rm Im} ( G_1 G_2^* ),  
\quad
    c_0 (\omega) = \frac{1}{3} a(\omega),                                        
\quad 
   c_4 (\omega) = 2 E_c^2  \vert G_1 \vert^2 , 
\nonumber\\
   c_5 (\omega) &=&2 E_c^2  \biggr [  \vert G_1 \vert^2 - y_m^2 \vert G_2\vert^2 
               +\omega^2 \vert G_1 - y_m G_2\vert^2 \biggr ],         
\quad
  c_6(\omega) =-2  E_c^2 \omega  \biggr [ \vert G_1 \vert^2 -  y_m {\rm Re }( G_1^* G_2)    \biggr ],  
\end{eqnarray} 
with $y_m =m_\Lambda/E_c$ and $E_c =M_{J/\psi}/2$.

The $P$- and $CP$- violating part of $R$ is given by the following nonzero functions:
\begin{eqnarray} 
  b_{1p}(\omega) &=& 2 E_c^2 \biggr (  -2 \omega \beta E_c {\rm Im } ( H_T G_1^*) + \omega\beta y_m  {\rm Re} ( F_A G_2^*) + 2 y_m d_J {\rm Re}  (G_1 G_2^*) \biggr ) ,   
\nonumber\\
  b_{1k} (\omega) &=&  2 E_c^2 \biggr ( -2 \beta E_c   {\rm Im}\biggr [ H_T ( \omega^2 G_1 
      + (1-\omega^2 ) y_m G_2)^*  \biggr ] +\beta  {\rm Re} \biggr [ F_A ((1+\omega^2) G_1 - \omega^2 y_m G_2 )^*\biggr ]
\nonumber\\      
   && + 2 \omega d_J \biggr ( \vert G_1\vert^2 + y_m {\rm Re} (G_1 G_2^* ) \biggr ) \biggr ) ,   
\nonumber\\
   c_1 (\omega ) &=& -4 \omega \beta E_c^3  {\rm Re} (H_T G_1^*), 
\quad 
  c_2(\omega)  = 4 \beta  E_c^3  {\rm Re}\biggr [ H_T ( \omega^2 G_1 
      + (1-\omega^2 ) y_m G_2)^*   \biggr ], 
\nonumber\\
   c_7 (\omega) &=& -2\beta E_c^2  \vert \hatp\times \hatk\vert {\rm Im} \biggr ( F_A G_1^* \biggr ), 
\nonumber\\
  c_8 (\omega ) &=& 2 E_c^2  \vert \hatp\times \hatk\vert \biggr ( \beta \omega E_c^2   {\rm Im} \biggr [ F_A  ( G_1-y_m G_2 )^* \biggr ] + 2 y_m d_J  {\rm Im} ( G_1^* G_2) \biggr ),                
\end{eqnarray}
with $\beta = \sqrt{ 1- y_m^2}$. $b_{2p}(\omega)$ and $b_{2k}(\omega)$ are obtained from $b_{1p}(\omega)$ and $b_{1k}(\omega)$ by replacing $H_T$ with $-H_T$, respectively.  It is interesting to note that $c_3 (\omega)$ is zero. Therefore, in the general 
form of the angular distribution one can discard the contribution from $c_3 (\omega)$. The results  in this and the last section can be extended to the case that the production is in continuum regions.

\par\vskip10pt
\noindent 
{\bf 4. Integrated Observables and Asymmetries} 

\par\vskip5pt

The various functions of $\omega$ charactering the general angular distribution in Eqs.(\ref{DECR}) and (\ref{GAD}) 
can be projected out by studying the angular distributions of $\hatl_p$ and $\hatl_{\bar p}$ in the space 
spanned by $\hatp$, $\hatk$ and $\hatn$. Because of the structure of $\rho$ in Eq.(\ref{RHOC}),  these functions of $\omega$ can only be one of the three forms  which are a constant,  a constant multiplied with $\omega$ or $\omega^2$.
Since the effects of symmetry-violation are expected to be small, it will be convenient to use integrated observables or asymmetries to directly measure the effects.  

The integrated observables can be built with the four unit vectors $\hatl_p$, $\hatl_{\bar p}$, 
$\hatp$ and $\hatk$. For any observable ${\mathcal O}$, its average which can be measured,  is  
 given by:
\begin{equation} 
    \langle {\mathcal O }\rangle = \frac{1}{{\mathcal N}}\int\frac{ d\Omega_k d\Omega_p d\Omega_{\bar p}} { (4\pi)^3}
    {\mathcal O} {\mathcal W} (\Omega) , 
\end{equation} 
with the normalization factor ${\mathcal N}$:
\begin{equation}     
  \quad  {\mathcal N} = \int\frac{ d\Omega_k d\Omega_p d\Omega_{\bar p}} { (4\pi)^3}    {\mathcal W}(\Omega)  = \frac{1}{3} E_c^2 \biggr [ 4 (\vert G_1 \vert^2 + y_m^2 \vert G_2\vert^2) +  ( \vert G_1 \vert^2 - y_m^2 \vert G_2\vert^2 )\biggr ] .  
\end{equation} 
In the following, we will only consider these observables whose nonzero values indicate $P$- or $CP$- violation.  

\par 
We first consider a set of observables in which only $\hatl_p$ is involved. 
These observables and their results are:    
\begin{eqnarray} 
  \langle \hatl_p \cdot \hatp\rangle &=&  
       \frac{4 \alpha }{9{\mathcal N}}  E_c^2 d_J \biggr ( 4 y_m {\rm Re} (G_1 G_2^*)+ 
        \vert G_1\vert^2  \biggr ), 
\nonumber\\
\langle \hatl_p\cdot\hatk\rangle  &=& \frac{4\alpha \beta }{9{\mathcal N}} E_c^2 \biggr [   2  {\rm Re} \biggr ( F_A G_1^*\biggr )-E_c {\rm Im} \biggr ( H_T G_1^* + y_m H_T G_2^* \biggr )\biggr ], 
\label{PL} 
 \end{eqnarray}
 These observables receive contributions only from $b_{1p}(\omega)$ and  $b_{1k}(\omega)$.  
 If any of the above observables is nonzero, it implies $P$ violation.    
 It is noted that the first observable is only sensitive to the parity-violation in the $J/\psi$ production. It is proportional to $d_J$. The second observable is only sensitive 
 to the parity violation in the decay of $J/\psi\to \Lambda\bar\Lambda$. Therefore, using these two observables one can distinguish where parity is violated, if the effect of parity violation is observed. 
 
Another set of observables we consider involve both of  $\hatl_p$ and $\hatl_{\bar p}$. This type of observables 
directly measures the quantum entanglement of spins in the process, or the matrix $C^{ij}$ in Eq.(\ref{DECR}). We introduce the following  observables as:
\begin{eqnarray}
 \langle \hatl_p \cdot \hatk \hatl_{\bar p} \cdot \hatn +\hatl_p \cdot \hatn  \hatl_{\bar p} \cdot \hat k \rangle 
   &=& -\frac{2\alpha\bar\alpha \pi }{9 {\mathcal N}}  y_m E_c^2 d_J {\rm Im}( G_1^* G_2), 
\nonumber\\
   \langle \hatl_p \cdot \hatp \hatl_{\bar p} \cdot \hatn +\hatl_p \cdot \hatn  \hatl_{\bar p} \cdot \hatp \rangle 
   &=& \frac{\alpha\bar\alpha \pi }{36  {\mathcal N}} E_c^2 \beta {\rm Im} \biggr [ F_A \biggr ( 3 G_1^*
      + y_m G_2^* \biggr ) \biggr ].  
      \label{PPL}
\end{eqnarray}
These two observables only receive contributions from $c_{7,8}(\omega)$. The first observable measures 
only the $P$ violation in the production of $J/\psi$, while the second measures the $P$ violation in the decay. 

The observables discussed in the above can be used for testing $P$- symmetry. To test CP symmetry one can use the second observable in Eq.(\ref{PL}) and that of $\bar \Lambda$   to check the following sum:
\begin{equation} 
 \langle \hatl_p\cdot\hatk\rangle + \langle \hatl_{\bar p} \cdot\hatk\rangle 
    =   - \frac{8\alpha \beta }{9{\mathcal N}} E_c^3 {\rm Im} \biggr ( H_T G_1^* + y_m H_T G_2^* \biggr )
        + {\mathcal O}(\bar\alpha -\alpha).  
\label{CPL1}   
\end{equation}  
If $CP$ is conserved in  in the decay of $\Lambda$ and $\bar\Lambda$, i.e., 
$\alpha=\bar \alpha$, $CP$-violation in the sum is only from $J/\psi$ decays. The following observable 
will only be sensitive to $CP$-violation in the decay of $J/\psi\to \Lambda\bar\Lambda$: 
\begin{equation} 
\langle (\hatl_p\times\hatl_{\bar p} )\cdot \hatk \rangle = -\frac{8\alpha\bar\alpha}{27 {\mathcal N}} \beta  y_m E_c^3 {\rm Re} 
  ( H_T G_2^* ). 
\label{CPL}   
\end{equation} 
This observable only receives contribution from $c_{1,2}(\omega)$.  It is noted that $\langle (\hatl_p\times\hatl_{\bar p} )\cdot \hatp \rangle =0$ after integration over $\Omega_k$.        
 
One can use any observables in Eqs.(\ref{PL}, \ref{PPL}, \ref{CPL1},\ref{CPL}) for probing $P$- or $CP$- violating effects.  For these observable ${\mathcal O}$ one can always define corresponding  asymmetries in events as:
\begin{equation} 
 {\mathcal A}({\mathcal O}  ) = \frac{ {\mathcal N}_{\rm event}  ({\mathcal O}>0) - {\mathcal N}_{\rm event} ({\mathcal O} < 0) } 
    { {\mathcal N}_{\rm event} ({\mathcal O} >0) +  {\mathcal N}_{\rm event} ({\mathcal O} < 0) }
      = \frac{1}{{\mathcal N}}\int\frac{ d\Omega_k d\Omega_p d\Omega_{\bar p}} { (4\pi)^3}
     \biggr ( \theta ({\mathcal O})- \theta (-{\mathcal O}) \biggr )  {\mathcal W} (\Omega) .  
\end{equation}   
We introduce the following asymmetries and give their results:  
\begin{eqnarray} 
&&\mathcal{A}_{d_J}={\mathcal A}(\hatl_p\cdot \hatp ) = \frac{3}{2}  \langle \hatl_p \cdot \hatp\rangle, \quad \mathcal{A}_{F_A}={\mathcal A}(\hatl_p\cdot \hatk - \hatl_{\bar p} \cdot\hatk) = \frac{3}{2}  \langle \hatl_p\cdot\hatk - \hatl_{\bar p} \cdot\hatk\rangle, 
\nonumber\\ 
 &&\mathcal{A}_{H_T}={\mathcal A}( (\hatl_p\times\hatl_{\bar p})\cdot \hat k ) =  \frac{3}{2}  \langle (\hatl_p\times\hatl_{\bar p})\cdot \hatk \rangle.    
 \end{eqnarray} 
 The statistical error of any asymmetry is given by $1/\sqrt{ N_{event}}$ with $N_{event}$ as the number of events.

 \par\vskip10pt
\noindent
{\bf 5. Numerical Results and Discussion}

\par\vskip5pt

 \par
The four form factors $G_1$, $G_2$, $F_A$ and $H_T$ in the decay introduced in Eq.(\ref{AMP}) are in general complex. One of them can be taken as real and positive because the overall phase has no effect.  
We will give our numerical results under the assumption that $G_1$ is real and positive, and 
$F_A$ and $H_T$ are real. With this assumption we have:  
\begin{eqnarray} 
 \mathcal{A}_{d_J}\approx 0.60 d_J, \quad \mathcal{A}_{F_A} \approx  606 F_A, \quad 
\mathcal{A}_{H_T}  \approx -46 (m_\Lambda H_T).
\end{eqnarray} 
In the above we have used the central values of $G_{1} \approx 1.14\times 10^{-3}$, $|G_2/G_1|^2 \approx 0.686$, the phase difference $\Delta \Phi = \Phi_{G_2} - \Phi_{G_1}=0.752$ and the averaged $\Lambda$ polarization $\alpha = 0.742$ extracted from  BESIII data~\cite{BESIII:2022qax}.

Non-zero values of $d_J$ or $F_A$ imply $P$ violation, and a non-zero $H_T$ implies $CP$ violation.  therefore experimental measurements of ${\mathcal A}_{d_J}$, ${\mathcal A}_{F_A}$ and ${\mathcal A}_{H_T}$ will provide new information about $P$- and $CP$- symmetries at the resonant production of $e^+e^-\to J/\psi$ followed by the subsequent decay of $J/\psi \to \Lambda \bar \Lambda$.  
Since nonperturbative effects are dominant in the process,  
it is in general difficult to estimate the contributions to these parameters from SM or new physics beyond SM,  where interactions violate $P$ and $CP$ are given at fundamental quark level.  E.g., if these symmetries in the formation of $J/\psi$ from a $c\bar c$-pair are violated, the produced $J/\psi$ is no longer a $P$- or $CP$- eigenstate, but a state mixed with states 
of different quantum numbers. The effects due to the mixed state are contained in $F_A$ and $H_T$. 
It is noted that $d_J$ is not affected by symmetry-violations in the formation. The leading contribution to $d_J$ in SM comes from $Z$-boson exchange, where the $e^+e^-$ pair annihilates into a virtual $Z$ and the $Z$ decays into a $c\bar c$ pair. 
We obtain:
\begin{equation}
     d_J = \frac{ 3- 8 \sin^2\theta_W }{32 \cos^2\theta_W \sin^2\theta_W} \frac{M^2_{J/\psi}}{M_Z^2} \approx 2.53\times 10^{-4}, 
\label{DJSM}      
\end{equation}       
where the on-shell scheme value\cite{PDG} of $0.223$ for $\sin^2\theta_W$ is used. 

\par 
Similarly, $F_A$ in SM also receives a nonzero contributions from $Z$-boson exchange in SM, where $J/\psi$ 
decays into $\Lambda\bar\Lambda$ through the $Z$-boson  exchange between $c \bar c$ and light quark pairs $q \bar q$.  Its contribution to the decay amplitude can be written: 
\begin{eqnarray}
 {\mathcal T}_Z  = - {g^2 g_V^c\over 4m^2_Z \cos^2\theta_W} \langle 0 \vert \bar c \gamma^\mu c \vert J/\psi \rangle  \langle  \bar\Lambda \Lambda \vert     g_A^q \bar q\gamma_\mu \gamma_5 q \vert  0 \rangle \;.
\end{eqnarray}
The matrix element of the charm quark current is given by  $ \langle 0 \vert \bar c \gamma_\mu c \vert J/\psi> = g_V  \epsilon_\mu$ with $g_V = 1.25 GeV^2$\cite{HMM}.  The sum over light flavors $q=(u,d,s)$ is implied with $g^c_V = 1/2-4\sin^2\theta_W/3$ and 
$g^d_A=g^s_A=-g^u_V=-1/2$. To give an estimate of ${\mathcal T}_Z$ or $F_A$ we relate the matrix element $ \langle  \bar\Lambda \Lambda \vert      \bar q\gamma_\mu \gamma_5 q \vert  0 \rangle$ to $\langle  \bar\Lambda  \vert      \bar q\gamma_\mu \gamma_5 q \vert  \Lambda \rangle$ by using crossing symmetry.  Neglecting the momentum difference 
between $\Lambda$ and $\bar\Lambda$, the sum $ \langle  \bar\Lambda  \vert    g_A^q  \bar q\gamma_\mu \gamma_5 q \vert  \Lambda \rangle$ is then given by  $- D/3 \bar \Lambda \gamma_\mu \gamma_5 \Lambda$ with $D = 0.80$ 
in the light quark flavor $SU(3)$ limit\cite{D-term}. Under the approximation we have the estimation: 
\begin{eqnarray}
F_A  \approx - {1\over 6} D g_V  {g^2\over 4 \cos^2\theta_W} {(1-8\sin^2\theta_W/3) \over m^2_Z} \approx - 1.07\times 10^{-6}\;.
\end{eqnarray}
Unlike $d_J$, there are other possible contributions to $F_A$ which we neglected in the above estimation. 

The $CP$-violating $H_T$ can only be generated  in SM at three loops or higher and is extremely small. When going beyond SM, a sizable $H_T$ can be generated. For example, a non-zero $\Lambda$ electric dipole moment $d_\Lambda$ induces a non-zero $H_T$ which has been calculated in \cite{HMM}:
\begin{eqnarray}
H_T = {2e \over 3 m^2_{J/\psi}}  g_V d_\Lambda\;.
\end{eqnarray} 
It should be kept in mind that form factors have a $q^2$-dependence in general. In our case $q^2$ is given by $q=M_{J/\psi}$. We neglect 
this dependence. Then the form factor $H_T$ is in our approximation related to $d_{\Lambda}$ which is in fact defined at $q^2=0$.   
The best experiment upper bound of $d_\Lambda$ is\cite{PDG} $1.5\times 10^{-16}$ ecm at the 95\% C.L.. Taking $d_\Lambda$ as the source for $H_T$, we obtain the upper bound for $H_T$ to be  
\begin{eqnarray}
m_\Lambda H_T = 2.23\times 10^{-4}\;.
\end{eqnarray}
Using the above estimated $d_J$, $F_A$ in the SM and the upper bound of $H_T$, we obtain the following theoretical predictions:
\begin{eqnarray}
&&\mathcal{A}_{d_J} \approx 1.52\times 10^{-4}\;,\;\;\mathcal{A}_{F_A} \approx - 6.51\times 10^{-4}\;,\;\;\mathcal{A}_{H_T} \le - 1.02\times 10^{-2}\;.
\end{eqnarray}
It should noted that the last two numerical results should be taken as order of magnitude estimates, because they are obtained under the assumption of relative phases of form factors and approximations mentioned before in this section.

\par 
For an asymmetry ${\mathcal A}$  its statistical error or sensitivity is given by $\delta {\mathcal A} = 
1/\sqrt{{\mathcal N}_{\rm event}}$, where ${\mathcal N}_{\rm event}$ is the number of events used to measure ${\mathcal A}$. In experiment there are other sources of errors beside statistical error. It is clear 
that an asymmetry can not be measured if its value is smaller than its statistical error.  With a given event number, one can estimate how well these asymmetries in the above or the parameters in the asymmetries in   
Eq.(29) can be measured. 
Currently, BESIII has collected $4.6\times 10^{6}$ events of the studied process\cite{BESIII:2022qax} which implies a statistic sensitivity of 
about $ 4.6 \times 10^{-4}$, the data can already be used to probe $P$-violation due to $F_A$ and to improve the bound on $d_\Lambda$ by a factor of a few. 
At the proposed STCF\cite{STCFC,STCFR}, where the proposed luminosity will be enhanced to $0.5\times 10^{35}cm^{-2}s^{-1}$,   $3.4\times 10^{12}$ $J/\psi$ with one year's running can be produced. 
It is expected that there will be about $6.4\times 10^{9}$ $e^+e^-\to J/\psi \to \Lambda \bar \Lambda$ data sample for analysis. The statistical error of an asymmetry will be reduced to $1.2 \times 10^{-5}$. The $P$-violating effects due to non-zero $d_J$ and $F_A$ can be detected to a high precision, and the bound on $\Lambda$'s edm will be improved by an order of magnitude or better. 

When going beyond the SM, the size of $d_J$ and $F_A$ may be larger. 
For example, if there is a leptoquark $S_1: (\bar 3, 3)(-5/6)$ couple to e and c, $\lambda_{ec} \bar L^c_L Q_L S_1$, a new ontribution to $d_J$ will be generated. The strongest constrain on the coupling $\lambda_{ec}$ come from electron
$a_e=(g_e-2)/2$ by exchange $c$ and $S_1$ in the loop. Using $\Delta a_e = a^{exp}_e - a^{SM}_e = (48\pm 30)\times 10^{-14}$ from data\cite{e-edm}, we find that within the allowed error range, $\Delta d_J$ can be an order of magnitude larger than the SM predictions. 
Exchange of a di-quark scalar $S_D: (3,1)(-4/3)$ may enhance $F_A$. $S_D$ can have Yukawa coupling to quarks, $\lambda_{uc} \bar u^c_R c_R S_D$. Exchange $S_D$ at tree level will produce an effective $F_A \sim 1.4\times 10^{-6} |\lambda_{uc}|^2\left ({100\mbox{GeV}\over m_{S_D}}\right )^2$. If the coupling $\lambda_{uc}(100\mbox{GeV}/m_{S_D})$ is not too much smaller than 1, there may be some enhancement compared with SM value. This may be possible if $\lambda_{uc}$ is the only non-zero coupling and close to its unitarity bound of $4\pi$ or so.

\par
We have only considered the asymmetries for testing discrete symmetries in the above. For an asymmetry the statistical error is simply determined by number of events. One can also consider the corresponding observables introduced in the last section. 
The statistical error of an ${\mathcal O}$ is given by: 
\begin{equation} 
   \delta \langle {\mathcal O} \rangle  = \sqrt { \frac{ \langle {\mathcal O}^2 \rangle  -\langle {\mathcal O} \rangle^2 }{{\mathcal N}_{\rm event}}     }    \label{error}
   \end{equation}       
where ${\mathcal N}_{\rm event}$ is the number of events. We have also considered these integrated  observables. Their statistical sensitivities are roughly at the same orders as the asymmetries discussed in the above. However, from the integrated observables it is possible to enhance statistical sensitivities through constructing the so-called observables so that the value of $\delta {\langle {\mathcal O} \rangle}/\langle {\mathcal O} \rangle$ is minimized\cite{DN}. To perform such a study it is better 
to include errors of other sources to have a realistic estimate of sensitivities.  

\par\vskip10pt
\noindent 
{\bf 6. Summary} 

\par\vskip5pt

We have studied how to test discrete $P$- and $CP$- space-time symmetries in the process $e^+ e^- \to J/\psi \to \Lambda\bar\Lambda$, where the polarization of the initial state is unobserved and that of $\Lambda$ and $\bar \Lambda$ is measured through their weak decays. 
The general angular distribution is derived for the chain process. The constraints on the distribution from $C$-,  $P$- and $CP$-symmetry are given.  From a general parametrization of the decay amplitude of $J/\psi$, we derive each component in the distribution.  We find that effects due to $P$-violation  in the $J/\psi$-production through  $e^+e^-$ annihilation and $J/\psi$ decays into $\Lambda \bar \Lambda$ can be measured separately with integrated observables proposed in this work.  The effects in $P$- violating observables are estimated by considering $Z$-boson exchange in SM.  
$CP$-violating observables are predicted in term of electric dipole moment of $\Lambda$. Our results show that  
 BESIII has already reached the sensitivity to probe the estimated $P$-violating effects. At the proposed STCF's they can be probed with high precision. $CP$- violating effect in $J/\psi \to \Lambda \bar \Lambda$ can also be tested. In particular, BESIII data already reached the level to improve $CP$- the current upper limit of  $\Lambda$'s electric dipole moment and the limit can be improved by an order of magnitude at the proposed STCF. We urge our experimental colleagues at BESIII to carry out measurements of observables  proposed here.

\par\vskip60pt

\noindent
{\bf Acknowledgments}
\par 
 We thank Dr. H.B. Li for very useful discussions.   
The work of J.P. Ma was supported by National Natural Science Foundation of P.R. China(No.12075299, 11821505, 11847612 and 11935017)   and by the Strategic Priority Research Program of Chinese Academy of Sciences, Grant No. XDB34000000. The work of 
X.G. He was supported by the Fundamental Research Funds for the Central Universities, by National Natural Science Foundation of P.R. China(No.12090064, 11735010, and 11985149) and by MOST 109–2112-M-002–017-MY3.
 
 \par\vskip10pt

 \par\vskip40pt

\par 


\vskip40pt
\begin{thebibliography}{99}  

\bibitem{PDG}
R.L. Workman {\it et al.} (Particle Data Group), Prog. Theor. Exp. Phys. 2022, 083C01 (2022).

\bibitem{BESIII:2022qax}
M.~Ablikim \textit{et al.} [BESIII],
Phys. Rev. Lett. \textbf{129}, no.13, 131801 (2022),
[arXiv:2204.11058 [hep-ex].

\bibitem{BESIII:2018cnd}
M.~Ablikim \textit{et al.} [BESIII],
Nature Phys. \textbf{15}, 631-634 (2019), 
arXiv:1808.08917 [hep-ex].

\bibitem{BESIII:2021ypr}
M.~Ablikim \textit{et al.} [BESIII],
Nature \textbf{606}, no.7912, 64-69 (2022), 
arXiv:2105.11155 [hep-ex].

\bibitem{STCFC} H.-P. Peng, “High Intensity Electron Positron Accelerator (HIEPA), Super Tau Charm Facility (STCF) in China”, talk at Charm2018, Novosibirsk, Russia, May
21 - 25, 2018.

\bibitem{STCFR} A.E. Bondar et al. (Super Charm-Tau Factory Collaboration), Project of a Super Charm-Tau
factory at the Budker Institute of Nuclear Physics in Novosibirsk,  Phys. Atom. Nucl. 76 (2013)
1072.

\bibitem{CPZ} W. Bernreuther, U. L\"ow, J.P. Ma and O. Nachtmann, Z. Phys. C43 (1989) 117.   

\bibitem{LeeYang} T.D. Lee and C.N. Yang, Phys. Rev. {\bf 108} (1957) 1645.  

\bibitem{BGR} A. Bondar {\it et. al.} JHEP 03 (2020) 076,  arXiv:1912.09760 [hep-ph]. 

\bibitem{HMM} X.G. He, J.P. Ma and B.H.J. Mckellar,   Phys.Rev. D47 (1993) 1744, 	arXiv:hep-ph/9211276. 

\bibitem{LiMa} H.B. Li and X.-X. Ma, Phys. Rev. D100, (2019) 076007, arXiv:1907.01151. 

\bibitem{FK} G. F\"aldt and A. Kupsc,  Phys. Lett. B772 (2017) 16–20, arXiv:1702.07288 [hep-ph].

\bibitem{D-term}
X.~G.~He, J.~Tandean and G.~Valencia,
Phys. Rev. D \textbf{72}, 074003 (2005)
[arXiv:hep-ph/0506067 [hep-ph]].
    
    
\bibitem{e-edm}
L.Morel, Z.Yao,P.Clade, and S. Guellati-Khelifa, Nature 588, 61 (2020).

\bibitem{DN} M.  Diehl and O. Nachtmann,  Z. Phys. C62 (1994) 397. 


\end{thebibliography}
\end{document}